\documentclass[onecolumn,prd,superscriptaddress,nofootinbib,showpacs,showkeys]{revtex4}
\usepackage{epsfig}
\usepackage{amsmath}

\newlength{\dinwidth}
\newlength{\dinmargin}
\newcommand{\NIMA}[3] {Nucl.\ Instr.\ and Meth.\ \textbf{A#1} (#2) #3}

\newcommand{\Zzero}{\mbox{${\mathrm{Z}}$}}

\newcommand{\qq}{\mbox{$\mathrm{q\overline{q}}$}}
\newcommand{\uu}{\mbox{$\mathrm{u\overline{u}}$}}
\newcommand{\qqp}{\mbox{$\mathrm{q'\overline{q}}$}}

\newcommand{\EGeV}{\mbox{$E(\mathrm{GeV})$}}
\newcommand{\EjjGeV}{\mbox{$E_{jj}(\mathrm{GeV})$}}
\newcommand{\Ej}{\mbox{$E_{j}$}}

\newcommand{\thetaqq}{\mbox{$\theta_{\mathrm{q\overline{q}}}$}}
\newcommand{\cosqq}{\mbox{$\cos\thetaqq$}}

\newcommand{\GeV}{\mbox{$\mathrm{GeV}$}}

\newcommand{\roots}{\mbox{$\sqrt{s}$}}

\newcommand{\rms}{\mbox{${\mathrm{rms}}_{90}$}}

\newcommand{\MARLIN}{\mbox{\sc Marlin}}
\newcommand{\MOKKA}{\mbox{\sc Mokka}}
\newcommand{\PANDORAPFA}{\mbox{\sc PandoraPFA}}
\newcommand{\GEANT}{\mbox{GEANT4}}

\newcommand{\LCIO}{\mbox{\sc Lcio}}

\newcommand{\Cambridge}{Dept. of Physics, Cavendish Laboratory, Univ. of Cambridge, JJ Thomson Av., Cambridge CB3 0HE, UK}
\def\etal{\mbox{{\it et al.}}}

\begin{document}
\begin{flushright}
  CU-HEP-07/08 
\end{flushright}

\title{ Progress with Particle Flow Calorimetry\footnote{To appear in Proceedings of LCWS07, DESY, Germany, June 2007.}} 
\author{Mark Thomson}
\affiliation{\Cambridge}

\begin{abstract}
One of the most important requirements for a detector at the
ILC is good jet energy resolution. It is widely believed 
that the particle flow approach to calorimetry is the key to 
achieving the ILC goal of a di-jet invariant mass resolution 
$\sigma_m/m < \Gamma_Z/m_Z$. This paper describes the current 
performance of the \PANDORAPFA\ particle flow algorithm. 
For simulated light quark jets in the Tesla TDR detector, 
the jet energy resolution achieved is better than
$\sigma_E/E \approx 3.4\,\%$ for jet energies in the range
$45-250$\,GeV. This represents the first demonstration that 
Particle Flow Calorimetry can reach the ILC jet energy resolution 
goals.
\end{abstract}

\maketitle

\section{Introduction}

Many of the interesting physics processes at the ILC will be characterised by
multi-jet final states, often accompanied by charged leptons 
and/or missing transverse energy associated with neutrinos or the 
lightest super-symmetric particles. The reconstruction of the 
invariant masses of two or more jets will provide a powerful tool 
for event reconstruction and identification. Unlike at LEP, where kinematic
fitting\cite{bib:mwfit} enabled precise jet-jet invariant mass reconstruction 
almost independent of the jet energy resolution, at the ILC this mass 
reconstruction will rely on the detector having excellent jet
energy resolution. The ILC goal is to achieve a 
mass resolution for $\mathrm{W}\rightarrow\qqp$ and 
$\mathrm{Z}\rightarrow\qq$ decays which is comparable to their natural 
widths, i.e. $\sigma_m/m = 2.7\,\% \approx \Gamma_W/m_W 
\approx \Gamma_Z/m_Z$. 
For a traditional calorimetric approach,
a jet energy resolution of $\sigma_E/E = \alpha/\sqrt{\EGeV}$ leads to a 
di-jet mass resolution of roughly 
$\sigma_m/m = \alpha/\sqrt{E_{jj}\mathrm{(GeV)}}$, where $E_{jj}$ 
is the energy of the di-jet system. At the ILC typical di-jet energies 
will be in the range $150-350$\,GeV, suggesting the goal of 
$\sigma_E/E \sim 0.3/\sqrt{\EGeV}$. This is more than a factor 
two better than the best jet energy resolution achieved at LEP, 
$\sigma_E/E = 0.6(1+|\cos\theta|)/\sqrt{E(\GeV)}$~\cite{bib:Aleph-jet}. 
Meeting the jet energy resolution goal is a major factor in the overall 
design of a detector for the ILC.

\section{The Particle Flow Approach to Calorimetry}

It is widely believed that the most promising strategy for achieving the ILC
jet energy goal is the particle flow analysis (PFA) approach to calorimetry. 
In contrast to a purely calorimetric measurement, PFA requires the 
reconstruction of the four-vectors of all visible particles in an event. The 
reconstructed jet energy is the sum of the energies of the individual 
particles. The momenta of charged particles are measured in the tracking 
detectors, while the energy measurements for photons and neutral hadrons 
are obtained from the calorimeters. The crucial step in PFA
is to assign the correct calorimeter hits to 
reconstructed particles, requiring efficient separation of nearby showers. 

Measurements of jet fragmentation at LEP have provided 
detailed information on the particle composition of jets 
(e.g.~\cite{bib:Knowles,bib:Green}).  On average, after the 
decay of short-lived particles, roughly 62\% of the energy of jets 
is carried by charged particles (mainly hadrons), around 27\% 
by photons, about 10\% by long-lived neutral 
hadrons ({\em e.g.} n/K$^0_{\mathrm{L}}$), and around 1.5\% by neutrinos.
Assuming calorimeter resolutions of $\sigma_E/E = 0.15/\sqrt{\EGeV}$
for photons and $\sigma_E/E = 0.55\sqrt{\EGeV}$ for hadrons, 
a jet energy resolution of $0.19/\sqrt{\EGeV}$ is obtained with the 
contributions from tracks, photons and neutral hadrons shown in 
Tab.~\ref{tab:res}. In practise it is not possible to reach this level 
of performance for two main reasons. Firstly, particles travelling at 
small angles to the beam axis will not be detected.
Secondly, and more importantly, it is not possible to perfectly 
associate all energy deposits with the correct particles. For example, 
if a photon is not resolved from a charged hadron shower, the photon 
energy is not counted. Similarly, if part 
of charged hadron shower is identified as a separate cluster the energy 
is effectively double-counted. This {\em confusion} degrades particle 
flow performance. Because confusion, rather than calorimetric performance, 
determines the overall performance, the jet energy resolution achieved 
will not, in general, be of the form  $\sigma_E/E = \alpha/\sqrt{\EGeV}$.

The crucial aspect of particle flow is the ability to 
correctly assign calorimeter energy deposits to the correct reconstructed 
particles. This places stringent requirements on the granularity of 
electromagnetic and hadron calorimeters. Consequently, particle flow 
performance is one of the main factors driving the overall ILC detector 
design. It should be noted that the jet energy resolution obtained for a 
particular detector concept is the combination of the intrinsic detector
performance and the performance of the PFA software. 

\begin{table}[th]
\renewcommand{\arraystretch}{1.2}
\begin{tabular}{l|cccc}
{Component}          & {Detector} & {Energy Fraction} & {Energy Res.} & {Jet Energy Res.} \\ \hline
Charged Particles ($X^\pm$) & Tracker           & $\sim0.6\,E_{\mathrm{jet}}$ & $10^{-4}\,E^2_{X^\pm}$  & 
   $<3.6\times10^{-5}\,E^2_{\mathrm{jet}}$ \\
Photons $(\gamma)$          & ECAL              & $\sim0.3\,E_{\mathrm{jet}}$ & $0.15\,\sqrt{E_\gamma}$ & 
   $0.08\,\sqrt{E_{\mathrm{jet}}}$ \\
Neutral Hadrons $(h^0)$     & HCAL              & $\sim0.1\,E_{\mathrm{jet}}$ & $0.55\,\sqrt{E_{h^0}}$  &
$0.17\,\sqrt{E_{\mathrm{jet}}}$ \\ 
\end{tabular}
\caption{Contributions from the different particle components to the jet-energy resolution
(all energies in GeV). The table lists the approximate fractions of charged particles, 
photons and neutral hadrons in a jet and the assumed single particle energy resolution. \label{tab:res}}
\renewcommand{\arraystretch}{1.0}
\end{table}

\section{The PandoraPFA Particle Flow Algorithm}

\PANDORAPFA\ is a {C++} implementation of a PFA 
algorithm running in the \MARLIN\cite{bib:marlin,bib:gaede} framework.
It was designed to be sufficiently generic for ILC detector 
optimisation studies and was developed and optimised using 
events generated with the \MOKKA\cite{bib:mokka} program, which
provides a \GEANT\cite{bib:geant4} simulation of the 
Tesla TDR\cite{bib:teslatdr} detector concept. The \PANDORAPFA\ algorithm
performs both calorimeter clustering and particle flow in 
eight main stages: 

\smallskip
\noindent
\noindent{\bf{i) Tracking:}} for the studies presented in this paper, the track {\em pattern recognition} 
               is
               performed using Monte Carlo information\cite{bib:marlin}. The track parameters are 
               extracted using a helical fit. The projections of tracks onto the 
               front face of the electromagnetic calorimeter are calculated using helical 
               fits (with no accounting for energy loss along the track).
               Neutral particle decays resulting in two charged particle tracks ($V^0$s) are
               identified by searching from pairs of non-vertex tracks which 
               are consistent with coming from a single point in the central tracking chamber. 
               Kinked tracks from charged particle decays to a single charged particle and a 
               number of neutrals are also identified, as are interactions in the tracking volume
               (prongs).
 
\smallskip
\noindent
{\bf{ii) Calorimeter Hit Selection and Ordering:}} isolated hits, defined on the basis of proximity 
          to other hits, are removed from the initial clustering stage. 
          The remaining hits are ordered into {\it pseudo-layers} which follow the detector 
          geometry so that particles propagating outward from the interaction region will 
          cross successive pseudo-layers. The assignment of hits to 
          pseudo-layers removes the dependence of the algorithm on the explicit detector 
          geometry whilst following the actual geometry as closely as possible. 
          Within each pseudo-layer hits are ordered by decreasing energy.

\smallskip
\noindent
{\bf{iii) Clustering:}} the main clustering algorithm is a cone-based forward projective method working 
                 from innermost to outermost pseudo-layer. In this manner hits are added 
                 to clusters or are used to seed new clusters. Throughout the 
                 clustering algorithm clusters are assigned a direction (or directions) 
                 in which they are growing. The algorithm starts by {\em seeding} clusters
                 using the projections of reconstructed tracks onto the front face of the
                 calorimeter. The initial direction of a track-seeded cluster is obtained 
                 from the track direction. The hits in each subsequent pseudo-layer are 
                 then looped over. Each hit, $i$, is compared to each clustered hit, $j$, 
                 in the previous layer. The vector displacement, ${\bf r_{ij}}$, is 
                 used to calculate the parallel and perpendicular displacement of the 
                 hit with respect to the unit vector(s) ${\bf\hat{u}}$ describing the
                 cluster propagation 
                 direction(s), $d_\parallel = {\bf r_{ij} .\hat{u}}$  and 
                 $d_\perp = |{\bf{r_{ij}\times\hat{u}}}|$. 
                 Associations are made using a cone-cut, 
                 $d_\perp < d_\parallel\tan\alpha + \beta D_{\mathrm{pad}}$, 
                 where $\alpha$ is the cone 
                 half-angle, $D_{\mathrm{pad}}$ is the size of a sensor pixel in the layer being 
                 considered, and $\beta$ is the number of pixels added to the cone radius. 
                 Different values of $\alpha$ 
                 and $\beta$ are used for the ECAL and HCAL with the default values set to
                 $\{\tan\alpha_{\mathrm{E}} = 0.3, \beta_{\mathrm{E}} =1.5\}$, 
                 and $\{\tan\alpha_{\mathrm{H}} = 0.5, 
                 \beta_{\mathrm{H}}=2.5\}$ respectively. Associations may be made with
                 hits in the previous 3 layers. If no association is made, the hit is used to 
                 seed a new cluster. This procedure is repeated sequentially for the hits in each 
                 pseudo-layer (working outward from ECAL front-face).

\smallskip
\noindent
{\bf{iv) Topological Cluster Merging:}} by design the initial clustering errs on the side of splitting up 
      true clusters rather than clustering energy deposits from more than one particle. 
      The next stage of the algorithm is to merge clusters from tracks and hadronic 
      showers which show clear topological signatures of being associated. 
      A number of track-like and shower-like topologies are searched for including looping 
      minimum ionising tracks, back-scattered tracks and showers associated with a 
      hadronic interaction. Before clusters are merged, a simple cut-based photon 
      identification procedure is applied. The cluster merging algorithms are only applied 
      to clusters which have not been identified as photons. 

\smallskip
\noindent
{\bf{v) Statistical Re-clustering:}} The previous four stages of the algorithm were found to 
      perform well for jets with energy less than $\sim50$\,GeV. 
      However, at higher energies the performance degrades rapidly due to
      the increasing overlap between hadronic showers from different particles. To address 
      this, temporary associations of tracks with reconstructed calorimeter clusters are made. 
      If the track momentum is incompatible with the energy of the associated 
      cluster re-clustering is performed. If $E_{\mathrm{CAL}}-E_{\mathrm{TRACK}} > 3.5\sigma_E$, 
      where $\sigma_E$ is the energy resolution of the cluster, 
      the clustering algorithm, described in {\em iii)} and {\em iv)} above, 
      is reapplied to the hits in that cluster.
      This is repeated, using successively smaller values of the $\alpha$s and $\beta$s in the 
      clustering finding algorithm (stage {\em iii)}) until the 
      cluster splits to give an acceptable track-cluster energy match. Similarly,
      if $E_{\mathrm{TRACK}}-E_{\mathrm{CAL}} > 3.5\sigma_E$ the algorithm attempts 
      to merge additional clusters with the cluster associated with the track. In doing
      so high energy clusters may be split as above.

\smallskip
\noindent
{\bf{vi) Photon Recovery and Identification:}} 
      A more sophisticated photon identification algorithm is then applied to the clusters.
      The longitudinal profile of the energy deposition, $\Delta E$, as a function of number of
      radiation lengths from the shower start, $t$, is compared to that expected 
      for an electromagnetic shower:
           $$ \Delta E \approx E_0 \frac{(t/2)^{a-1} e^{-t/2}} {\Gamma(a)} \Delta t \ \ \ \ \mathrm{where} \ \ \ a = 1.25 +\frac{1}{2}\ln\frac{E_0}{E_c},$$ 
      $E_0$ is the shower energy and $E_c$ is the critical energy which is tuned give the
      appropriate average shower profile in the ECAL.
      The resulting level of agreement is used to improve the tagging of photons and
      to recover primary photons merged with hadronic showers.

\smallskip
\noindent
{\bf{vii) Fragment Removal:}} At this stage there is still a significant number of  
      ``neutral clusters'' (not identified as photons) which are {\em fragments} 
       of charged particle hadronic showers. An attempt is made to
      identify these clusters and merge them with the appropriate parent cluster. 
      All non-photon neutral clusters, $i$, are compared to all clusters with associated tracks,
       $j$. 
      For each combination a quantity, $e_{ij}$, is defined which encapsulates the evidence 
      that cluster $i$ is a fragment from cluster $j$. The requirement, $R_{ij}$, for the 
      clusters to be merged,
      {\em i.e.} the cut on $e_{ij}$, depends on the location of the neutral cluster and
      the change in the  $\chi^2$ for the track$-$cluster energy consistency that would occur
      if the clusters were merged, $\Delta\chi^2 = (E_{\mathrm{TRACK}}-E_j)^2/\sigma^2_E
      - (E_{\mathrm{TRACK}}-E_i-E_j)^2/\sigma^2_E$. If $e_{ij} > R_{ij}$ the clusters are
      merged. This {\em ad hoc} procedure gives extra weight to potential 
      cluster matches where the consistency of the track momentum and associated cluster 
      energy improves
      as a result of the match.

\smallskip
\noindent
{\bf{viii) Formation of Particle Flow Objects:}} 
     The final stage of the algorithm is to create Particle Flow Objects (PFOs) from
     the results of the clustering. Tracks are matched to clusters on the basis 
     of the distance closest approach of the track projection into the first 10 layers 
     of the calorimeter. If a hit is found within 50\,mm of the track extrapolation an 
     association is made. 
     If an identified kink is consistent with being from a $K^\pm\rightarrow\mu^\pm\nu$ or
     $K^\pm\rightarrow\mu^\pm\nu$ decay the parent track is used to form the PFO. The 
     reconstructed PFOs are written out in \LCIO\cite{bib:marlin} 
     format.
 
\section{Current Performance}

\begin{figure}[hbtp]
\epsfxsize=1.0\columnwidth
\centerline{\includegraphics[width=0.95\columnwidth]{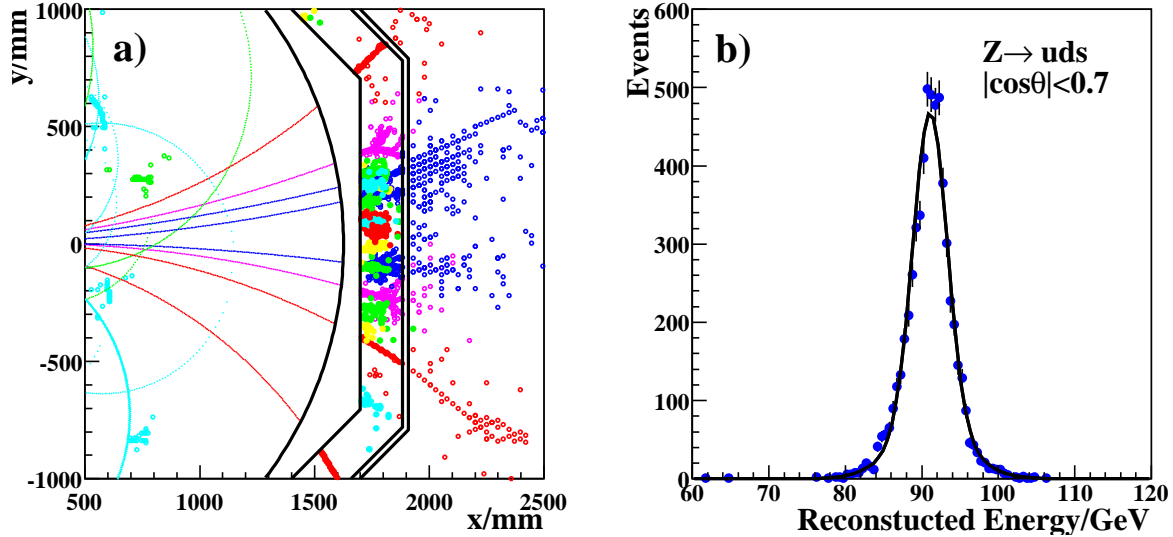}}
\caption{a) \PANDORAPFA\ reconstruction of a 100\,GeV jet in the \MOKKA\
             simulation of the Tesla TDR detector. 
             b) The total reconstructed energy from reconstructed 
             PFOs in $\Zzero\rightarrow{\mathrm{uds}}$ events for initial quark directions
             within the polar angle acceptance $|\cosqq|<0.7$. 
             The solid line shows a fit to two Gaussians with a common mean; the 
             broader Gaussian is constrained to contain 25\,\% of the events.
             The narrow Gaussian has a width of 2.2\,\GeV.}
\label{fig:figure1}
\end{figure}

\begin{center}
\begin{figure}[hbtp]
\epsfxsize=11.25cm
\centerline{\epsfbox{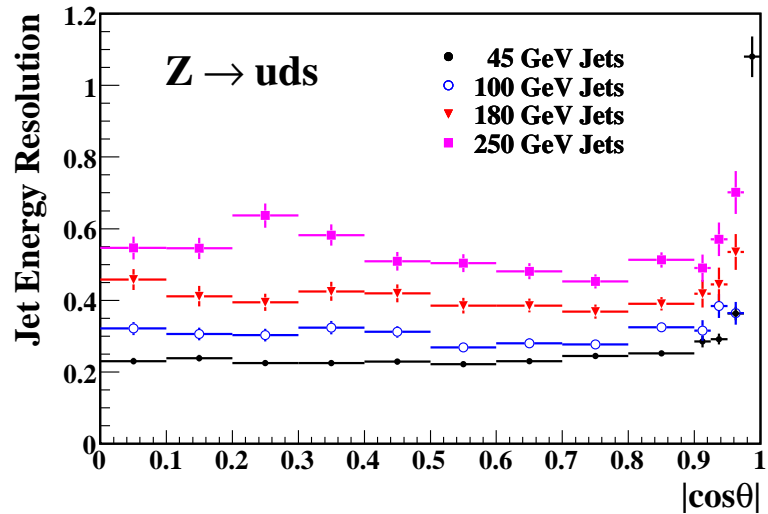}}
\caption{The jet energy resolution, defined as the $\alpha$ in $\sigma_E/E=\alpha\sqrt{\EGeV}$, 
        plotted versus $\cosqq$ for four different values of $\roots$.}
\label{fig:figure2}
\end{figure}
\end{center}

Fig.~\ref{fig:figure1}a) shows an example of a \PANDORAPFA\ reconstruction of 
a 100\,GeV jet from a $\Zzero\rightarrow\uu$ decay at $\roots=200$\,GeV. The ability
to track particles in the high granularity Tesla TDR calorimeter can be seen clearly. 
Fig.~1b) shows the total PFA reconstructed 
energy for $\Zzero\rightarrow{\mathrm{uds}}$ events with $|\cosqq|<0.7$, 
where $\thetaqq$ is the polar angle of the generated $\qq$ system. These events were
generated at $\roots=91.2$\,GeV using the Tesla TDR detector model with a HCAL consisting of
63 layers and in total 6.9 interaction lengths.
The root-mean-square deviation from the mean (rms) of the distribution is 2.8\,GeV. 
However, quoting the rms as a measure of the performance over-emphasises the 
importance of the tails. It is conventional to quote the performance in terms of
of $\rms$, which is defined as the rms in the smallest range of reconstructed energy which 
contains 90\,\% of the events. For the data shown in Fig.~\ref{fig:figure1}b) the resolution 
achieved is $\rms/E = 0.23/\sqrt{\EGeV}$, equivalent to a single jet energy resolution 
of 3.3\,\%. 
The majority of interesting ILC physics will consist of final states with
at least six fermions, setting a ``typical'' energy scale for ILC jets 
as approximately 85\,GeV and 170\,GeV at $\roots=500$\,GeV and $\roots=$1\,TeV respectively. 
Fig.~\ref{fig:figure2} shows the jet energy resolution for $\Zzero\rightarrow$uds events 
plotted against $|\cosqq|$ for four different values of $\roots$. The current performance 
is summarised in Tab.~\ref{tab:resvsE}. The observed jet 
energy resolution in simulated events is not described by the expression 
$\sigma_E/E = \alpha/\sqrt{\EGeV}$. This is not surprising, 
as the particle density increases it becomes harder to correctly
associate the calorimetric energy deposits to the particles and the confusion term increases.
The table also shows a measure of the single jet energy resolution, obtained by dividing $\rms$ by 
$\sqrt{2}$. For the jet energies considered ($45-250$\,GeV) the fractional energy 
resolution is significantly better than the ILC requirement of 3.8\,\% obtained from
the consideration of gauge boson di-jet mass resolution. It should be noted that
in a real physics analysis the performance is likely to be degraded by jet finding, jet-pairing
and the presence of missing energy from semi-leptonic heavy quark decays. Nevertheless the
results presented in this paper already provide a strong indication that Particle Flow Calorimetry
will be able to deliver the ILC jet energy goals and it is expected that the performance
of \PANDORAPFA\ will improve with future refinements to the algorithm.

\section{Conclusions}  

Particle flow calorimetry is widely believed to be the key to reaching the ILC jet 
energy resolution goal of a di-boson mass resolution of $\sigma_m/m < 2.7\,\%$. 
Consequently, the design and optimisation of detectors for the ILC depends both on 
hardware and on sophisticated software reconstruction. Based on the \PANDORAPFA\ 
reconstruction of simulated events in Tesla TDR detector concept, it has now been 
demonstrated that particle flow calorimetry can meet this goal at the ILC. This
was not true at the time of LCWS06 and, thus, represents a significant step forward
in the design and future optimisation of the ILC detector(s).

\begin{table}[ht]
\begin{center}
\begin{tabular}{r|rcc}
  Jet Energy        & $\rms$ &  $\rms/\sqrt{\EjjGeV}$ & $\rms/\sqrt{2}\Ej$  \\ \hline
  45 GeV            & 2.2\,GeV        &  23\,\%   &          3.3\,\%  \\
  100 GeV           & 4.1\,GeV        &  29\,\%   &          2.9\,\%  \\
  180 GeV           & 7.4\,GeV        &  39\,\%   &          2.9\,\%  \\
  250 GeV           & 12.0\,GeV       &  54\,\%   &          3.4\,\%  \\
\end{tabular}
\caption{Jet energy resolution for $\Zzero\rightarrow$uds events with $|\cosqq|<0.7$, 
            expressed as, $\rms$ for the di-jet energy distribution, the effective
            constant $\alpha$ in $\rms/E = \alpha(E_{jj})/\sqrt{\EjjGeV}$, and the
            fractional jet energy resolution for a single jet.
\label{tab:resvsE}}
\end{center}
\end{table}


\begin{footnotesize}

\end{footnotesize}


\begin{thebibliography}{99}
\bibitem{bib:mwfit} M.~A.~Thomson, Proc. of EPS-HEP 2003, Aachen.
Topical Vol. of Eur. Phys. J. C Direct (2004).
\bibitem{bib:Aleph-jet}ALEPH Collaboration, D.~Buskulic et al., Nucl.~Inst.~Meth. {\bf A360} (1995) 481.
\bibitem{bib:Knowles}I.G.~Knowles and G.D.~Lafferty, J.~Phys. {\bf G23} (1997) 731.
\bibitem{bib:Green} M.~G.~Green, S.~L.~Lloyd, P.~N.~Ratoff and D.~R.~Ward, ``Electron-Positron Physics at the Z'', IoP Publishing (1998).
\bibitem{bib:marlin} http://www-flc.desy.de/ilcsoft/ilcsoftware/.
\bibitem{bib:gaede} F.~Gaede,  ``Status of ILC-LDC Core Software'', Proceedings of LCWS07, DESY, June 2007.
\bibitem{bib:mokka} http://polywww.in2p3.fr/activites/physique/geant4/tesla/www/mokka/.
\bibitem{bib:geant4} GEANT4 collaboration, S. Agostinelli \etal, \NIMA{506}{2003}{3}; \\
                     GEANT4 collaboration, J. Allison \etal, IEEE Trans. Nucl. Sci. 53 (2006) 1.
\bibitem{bib:teslatdr} TESLA Technical Design Report, DESY 2001-011, ECFA 2001-2009 (2001).
\end{thebibliography}
\end{document}